\documentstyle[emulateapj]{article}

\begin{document}

\def\kms{km~s$^{-1}$}
\def\msun{$M_{\odot}$}
\def\rsun{$R_{\odot}$}
\def\lsun{$L_{\odot}$}
\def\halpha{H$\alpha$}
\def\Teff{T$_{\rm eff}$}
\def\logg{$log_g$} 

\tighten

\slugcomment{\bf}
\slugcomment {Ap.J. Letters, in press} 

\title{AN IMPROVED RED SPECTRUM OF THE METHANE OR T-DWARF
SDSS~1624+0029: ROLE OF THE ALKALI METALS}

\author{James Liebert\altaffilmark{1}, I. Neill Reid\altaffilmark{2}, 
Adam Burrows\altaffilmark{1}, Adam J. Burgasser\altaffilmark{3}, J. Davy 
Kirkpatrick\altaffilmark{4}, and John E. Gizis\altaffilmark{4}}

\altaffiltext{1}{Department of Astronomy and Steward Observatory,
The University of Arizona, Tucson AZ 85721, liebert@as.arizona.edu,
aburrows@as.arizona.edu}

\altaffiltext{2}{Department of Physics and Astronomy, University of
Pennsylvania, 209 South 33rd St., Philadelphia, PA 19104-6396,
inr@morales.physics.upenn.edu}

\altaffiltext{3}{Division of Physics, M/S 103-33, California Institute
of Technology, Pasadena CA 91125, diver@cco.caltech.edu}

\altaffiltext{4}{Infrared Processing and Analysis Center, California 
Institute of Technology, Pasadena, CA 91125, davy@ipac.caltech.edu,
gizis@ipac.caltech.edu}

\begin{abstract} A Keck~II low resolution spectrum shortward of
one micron is presented for SDSS~1624+0029, the first field methane or T
dwarf discovered in the Sloan Digital Sky Survey. Significant flux is
detected down to the spectrum's short wavelength limit of 6200\AA.  The
spectrum exhibits a broad absorption feature centered at 7700\AA, which
we interpret as the K~I 7665/7699 resonance doublet.  The observed flux
declines shortward of 7000\AA, due most likely to the red wing of the
Na~I doublet.  Both Cs~I doublet lines are detected more strongly than
in an earlier red spectrum.  Neither Li~I absorption nor
H$\alpha$ emission are detected.

An exploratory model fit to the spectrum suggests that the shape of the
red spectrum can be primarily accounted for by the broad wings of the
K~I and Na~I doublets.  This behavior is consistent with the argument
proffered by Burrows, Marley and Sharp that strong alkali absorption is
principally responsible for depressing T dwarf spectra shortward of
1$\mu$m.  In particular, there seems no compelling reason at this time
to introduce dust or an additional opacity source in the atmosphere of
the SDSS object.  The width of the K~I and strengths of the Cs~I lines
also indicate that the Sloan object is warmer than Gl~229B.

\end{abstract}

\keywords{stars: individual (SDSS1624+0029 -- stars: brown dwarfs --
stars: atmospheres}
\vfill
\eject

\section{INTRODUCTION}

Methane or T dwarfs are substellar objects cooler than L and M dwarfs,
and have near-infrared (1-2$\mu$m) spectra dominated by molecular
absorption due to water, methane, and pressure-induced molecular
hydrogen.  Methane is expected to remain an important atmospheric
constituent down to the temperature of Jupiter ($\sim$125~K), where it
also is prominent in the infrared spectrum.  The prototype of the class
is the companion to the nearby M dwarf star Gl~229 (Nakajima et
al. 1995, Oppenheimer et al.  1995).  Model atmosphere analyses fitting
synthetic spectra to detailed spectrophotometric and photometric
observations indicate a temperature for this object near or slightly
below 1,000~K (Marley et al. 1996, Allard et al. 1996).  This past year
has seen the discovery of several, similar field T dwarfs, found first
in the Sloan Digital Sky Survey (Strauss et al. 1999, S99; Tsvetanov et
al. 2000), shortly thereafter in the Two Micron All Sky Survey
(Burgasser et al. 1999) data sets, and also in an ESO survey (Cuby et
al. 1999).

All of the new objects also have 1--2$\mu$m spectra characterized by the
very strong molecular absorbers listed above.  Unfortunately, at least
at low spectral resolution, the differences among their spectra appear
somewhat subtle.  The current field surveys by SDSS and 2MASS are
magnitude limited, and therefore likely to identify the warmest (highest
luminosity) T dwarfs.  Still, the known parallax for Gl~570D shows that
this object must be substantially cooler than the prototype, yet its
infrared spectrum is similar to the others (Burgasser et al. 2000).  The
relative strengths of the molecular bands are not strongly dependent on
the effective temperature.  The molecular absorbers also are effective
in hiding the weaker atomic line transitions which might be useful
discriminants of temperature.  At least initially, it is proving
difficult to establish spectral types and a temperature sequence for the
T dwarfs at the wavelengths where they are easiest to observe.

It is possible to observe the brightest of the new T dwarfs at
wavelengths significantly shortward of 1$\mu$m, where the atmospheres may
prove to be more transparent.  Model calculations indicate that there
may be few molecular and atomic opacity sources, and those that are
present may be more sensitive to temperature.  In particular, the
behavior of the alkali resonance doublet features which reside generally
in this red part of the spectrum (0.5-0.9$\mu$m) could be particularly
useful in diagnosing the temperature and testing for the formation of
dust grains (Burrows and Sharp 1999; Lodders 1999; Tsuji, Ohnaka \& Aoki
1999; Burrows, Marley \& Sharp 2000; hereafter, BMS).  These papers
generally predict that different alkalis should precipitate out as
sulfides, salts or other condensates over a range of \Teff\ below
about 1,500~K, in the order (with decreasing \Teff) according to BMS of
Li first, then Cs, K and Na.  Indeed, the latter two alkali features are
the most prominent features in the red spectra of the somewhat-warmer
late L dwarfs (Kirkpatrick et al. 1999, Mart{\'{\i}}n et al. 1999, Reid
et al. 2000).

The red spectra of field T dwarfs can be relevant to another controversy
regarding the spectrum of Gl~229B: The companion object shows too little
red flux relative to the model predictions of Marley et al. (1996) and
Allard et al. (1996). These authors concluded that an additional opacity
source exists shortward of 1$\mu$m.  Golimowski et al. (1998) suggested
as the solution that TiO returned to gaseous form (it precipitates out
in M-L dwarfs above 2,000~K) in Gl~229B.  However, this hypothesis
predicts TiO band absorption at the wavelengths seen in M dwarfs, but
these are seen neither in late L dwarfs nor in Gl~229B.  Griffith et
al. (1998) turned to solar system physics for an intriguing answer: they
hypothesized a population of small photochemical haze particles
analogous to the red Titan Tholins (Khare and Sagan 1984), heated by
ultraviolet radiation from the primary M dwarf to temperatures at least
50\% higher than the \Teff.  Dust is invoked in two other proposed
solutions to the Gl~229B spectral slope: Tsuji, Ohnaka \& Aoki (1999)
describe a hybrid atmospheric model with a warm dust layer that
effectively blocks short-wavelength flux.  Pavlenko, Zapaterio Osorio \&
Rebolo (2000) attempt to fit the red spectrum with a scattering dust
opacity which increases sharply to shorter wavelengths.  Finally, BMS
suggest that the alkali opacity alone -- in particular the broad wings
of K and Na -- is the agent which depresses the emergent flux out to
1$\mu$m.  Provided K and Na still exist in atomic form at the relevant
temperature, they argued that there is no need to invoke dust or some
additional absorber at short wavelengths.

SDSS~1624+0029 (hereafter, SDSS~1624), the first field T dwarf, was
found in preliminary Sloan Digital Sky Survey data (S99).  The discovery
paper includes a red spectrum with the Apache Point 3.5-m telescope,
with detected flux down to 8000\AA.  The accessible Cs~I features at
8521\AA\ and 8943\AA\ appeared weak or absent, while both are distinct
features in the Gl~229B red spectrum (Oppenheimer et al. 1998).  The
apparent weakness of the Cs~I features and shallower red spectral slope
of SDSS~1624 led BMS to the preferred conclusion that this object ``is
tied to a lower core entropy,'' which would normally imply a lower
\Teff\ than that of Gl~229B.  In contrast, the Sloan source showed
somewhat shallower methane absorption in the infrared spectrum,
suggesting to Nakajima et al. (2000) that it is warmer than the
prototype.  The red spectrum also shows that a field object can show a
similar excess below 1$\mu$m as the companion object Gl~229B, thus
demonstrating that the excess does not depend upon the presence of a
nearby source of potential ultraviolet photons (ie., to produce the
``Titan Tholins'').

We discuss here a red spectrum of SDSS~1624 obtained with the Keck~II
LRIS spectrograph that extends the detection of flux down to 6200\AA,
and has improved signal-to-noise ratio at longer red wavelengths.  We
believe this observation allows us to test the roles of the alkali
metals and the need for dust and/or an additional short-wavelength
absorber.

\section{THE SPECTRUM AND ITS FEATURES}

Two consecutive spectra of SDSS~1624 were obtained on 1999 July 16 with
the Keck~II telescope and LRIS using the configuration described for
most observations in Kirkpatrick et al. (1999). Each had an exposure
time of 1800 seconds.  Reduction was done with standard IRAF tools.  The
averaged spectrum spanning the entire wavelength interval of
6,300--10,100\AA\ at 9\AA\ resolution is shown in Fig.~1.  Significant
flux is detected over this entire range, as shown in an inset, which
details the 6,300--8,200\AA\ spectrum on an appropriate vertical scale.
The variance spectrum from the standard IRAF task is also shown in the
short wavelength inset where the signal-to-noise ratio (SNR) is
smallest.  Longward of 8200\AA\, the variance spectrum (not shown) rises
slowly to about 4$\times$10$^{-18}$ near one micron, except for the
noisier intervals affected by the strongest atmospheric OH bands.  Some
significant conclusions may be drawn just from inspection of these
figures.

(1) The spectrum at the shorter wavelengths, not accessible in previous
observations, shows a strong, broad dip to zero flux centered precisely
on the 7700\AA\ blended doublet (see upper inset).  Note that the
variance plot remains flat over this interval, a strong indication that
the feature is not due to any change in the noise.  A ``pseudo''-EW
(pEW) of 390\AA\ for the line cores was measured over the interval
7300-8100\AA.  This was estimated from the IRAF ``splot'' routine which
simply fits a linear continuum across the stated interval.  It is
recognized that the flux levels at the interval boundaries are not the
true continuum.  Moreover, the procedure ignores the the considerable
absorption in the extended wings of this doublet.  Nonetheless, the pEW
estimate may serve as a useful benchmark of the K~I strength for quick
comparison with any similar spectra obtained for other objects.  The
strength of the feature confirms the suggestion of Tsuji et al. (1999)
and BMS that the red wing of this feature is a substantial absorber
shortward of one micron in this T dwarf; presumably the same feature
contributes to the flux deficiency in Gl~229B.  Indeed, it was already
recognized that this feature increases in strength with later types
among the L dwarfs (Kirkpatrick et al. 1999; Mart{\'{\i}}n et al. 1999).

(2) The detected flux rises shortward of 7700\AA, revealing the blue
wing of the K~I feature.  A broad maximum of the flux level may be
reached near 7000\AA, but significant flux is detected down to 6200\AA,
after moderate decline in the flux level shortward from 7000\AA.  This
last decline is likely due to the red wing of the Na~I resonance doublet
centered near 5892\AA.  Na is normally the most abundant of the five
alkalis in a Popuation~I mix.  Both Na and K are expected to survive to
lower temperatures than Li, Cs and Rb, so their presence here is not
surprising.  Indeed, subordinate lines of K~I were reported in the S99
(see also Nakajima et al. 2000) infrared spectrum.  The SNR dips below
unity at the last few hundred \AA, but the conclusions about the shape
of the continuum are robust.  No significant absorption features are
claimed.

(3) Significant Cs~I absorption lines, both members of the
well-separated doublet, are easily detected.  The pEWs of 6.5\AA\ and
6.1\AA, for the respective 8521\AA\ and 8943\AA\ transitions, were
measured over full width intervals of 25\AA\ each.  The Cs lines appear
much stronger than was apparent in the Arc 3.5-m spectrum of S99.  We
note that the apparent strength of the 8943\AA\ line could be enhanced
by contribution from an overlapping weak CH$_4$ band.  For the same
intervals, pEWs of 6.5\AA\ and 5.4\AA\ were measured for the web-posted
Oppenheimer et al. (1998) spectrum of Gl~229B.  While detailed modeling
of the alkali line profiles and red spectrum needs to be done, the
conclusion of BMS that SDSS~1624 may have a lower temperature than
Gl~229B needs to be reconsidered.  It is also possible, however, that
the Cs line strength varies with time and/or location on the surface.

(4) No significant H$\alpha$ emission or Li~I absorption is detected, to
limits we estimate as pEWs of 15\AA\, very approximately, since the SNR
is near unity.

(5) The $\sim$9300\AA\ H$_2$O band reported by S99 is strong in this
spectrum.  A possible absorption which appears to be strongest near
9955\AA, could be due at least in part to FeH, if the wavelength
calibration is poor near the edge of the spectrum.  A possible
absorption feature near 8343\AA\ may be H$_2$O; one near 8624\AA\ (see
lower inset of Fig.~1) coincides with both CH$_4$ and CrH bands, more
likely the former.

\section{AN EXPLORATORY MODEL FIT AND POSSIBLE IMPLICATIONS}

Several authors have explored quantitative fits to the alkali resonance
lines as a tool for estimating \Teff\ values for L and T dwarfs.
Unfortunately, the detailed treatment of the line broadening poses a
complicated problem -- see BMS Section 3.  Only a few key points are
mentioned here.  Available empirical data (cf. Nefedov, Sinel'shchikov,
and Usachev 1999) provide valuable clues to what might be the best
functional form for the profile.  However, broadening parameters remain
uncertain.  In any case, simple treatments (ie. Lorentzian profiles)
used previously in the literature appear inappropriate.

The detailed spectral fit depends upon several parameters, including the
\Teff, gravity, and abundances, but also the line profile shapes, and the
degree of rainout (Burrows and Sharp 1999; BMS).  We emphasize that the
physical treatment of these last two remain uncertain.  A detailed
exploration of these many parameters would be required, at minimum, to
achieve quantitative estimates of the \Teff, surface gravity and
abundances.  It is questionable whether a satisfactory, unique solution
will be found until a trigonometric parallax can help fix the luminosity
and radius.  What is shown below is an exploratory fit which we argue
nonetheless yields useful qualitative conclusions.  The techniques
employed are described more generally in BMS.

Figure 2 shows a comparison between a representative model spectrum
(dashed line) and a smoothed version of the SDSS~1624 spectrum
(solid).  Superposed is the Leggett et al. Gliese 229B spectrum of
Oppenheimer et al. (1998).  However, the observed flux has been
transformed to an absolute flux for an assumed distance to the Sloan
dwarf of 10 parsecs (S99), and is in milliJanskys, while the wavelength
is in microns.

It may be seen clearly that the K~I and Na~I doublets and their broad
wings dominate the spectrum.  For this illustrative model fit, we
assumed \Teff\ = 1100~K, gravity = $10^5$ cm s$^{-2}$, abundances of one
half solar, alkali line wing cutoff parameters defined in BMS of 0.2
(Na~I) and 0.5 (K~I), and an intermediate degree of rainout for the
alkalis (BMS).  The SDSS~1624 data were smoothed with a 10-\AA\ boxcar
function, which, among other things, muted the depth of the Cs lines
relative to the model, but are of similar strength.  No strong Li~I
absorption is predicted.

To obtain a reasonable fit, it is not clear to us that a dust component
or additional source of red opacity is required.  Tsuji's need for
additional red opacity may be explainable by (1) underestimation of the
alkali wing opacity due to the assumption of a Lorentzian, and (2) the
I-band broad band flux was plotted at the wrong mean wavelength (see
BMS).  Although the presence of dust in the atmosphere certainly cannot
be precluded, the alkalis appear to be the dominant cause of the unique
shape of the red energy distribution.  The detection of flux to the blue
boundary of the spectrum also has consequences.  In particular, a
Rayleigh scattering dust opacity, as suggested by Pavlenko et al (2000),
would have more than double the opacity at 7000\AA\ than at 8400\AA.
Finally, the observed narrowness of the 7700\AA\ feature (relative to
our models at the Gl~229B temperature near 950~K) and the presence of
strong cesium features together argue that the effective temperature of
SDSS~1624 is above that of Gliese 229B (BMS), in concurrence with
Nakajima et al. (2000).

We emphasize that no concerted attempt was made to find a rigorous fit,
that other combinations of parameters are still viable, and that, given
the SNR of the data at the shorter wavelengths, there are indeed
parameter degeneracies.

\acknowledgments { } This research is supported by a NASA JPL grant 
(961040NSF) permitting us to undertake a core science project on very
low mass objects discovered in the $2MASS$ survey.  AB acknowledges
support from NASA grants NAG5-7499 and NAG5-7073. The model curve was
computed based upon a temperature/pressure profile generated by
M. Marley (private communication) and the models in Burrows et
al. (1997).  We wish to acknowledge helpful suggestions from an
anonymous referee.

\newpage

\begin{figure}

\caption{The Keck~II spectrum of SDSS1624+0029, F$_\lambda$ vs. 
$\lambda$(\AA), boxcar-smoothed by 5 pixels (10\AA).  Two inset boxes
are discussed in the text.  The noise (variance) spectrum is shown 
in the top inset (shifted upward by 10$^{-18}$) and is flat over 
this interval.}

\end{figure}   

\begin{figure}

\caption{Comparison between a representative model spectrum (dashed) 
and a smoothed version of the SDSS1624 spectrum (solid), along with 
the published spectrum of Gl~229B (see text for details). The flux is in
milliJanskys, and the wavelength in microns. The normalizations of the
observations are based on an arbitrary distance assumed for SDSS1624 and
the trigonometric parallax of Gl~229B.}

\end{figure}   


\begin{references}
\overfullrule=0pt

\reference {} Burgasser, A., Kirkpatrick, J.D., Brown,
M.E., Reid, I.N., Gizis, J.E., Dahn, C.C., Monet, D.G., Beichman,
C.A., Liebert, J., Cutri, R.M., \& Skrutskie. M.F. 1999, \apjl, 522, L65

\reference {} Burgasser, A., Kirkpatrick, J.D., Cutri, R.C., McCallon,
H., Kopan, G., Gizis, J.E., Liebert, J., Reid, I.N., Brown, M.E., Monet,
D.G., Dahn, C.C., Beichman, C.A., \& Skrutskie. M.F. 2000, \apjl, in 
press. 

\reference {} Burrows, A., Marley, M., Hubbard, W.B., Lunine, J.I.,
Guillot, T., Saumon, D., Freedman, R., Sudarsky, D., \& Sharp, C.M.
1997, \apj, 491, 856 

\reference {} Burrows, A., Marley, M., \& Sharp, C. 2000, {\it Ap.J.}, 
531, in press (1 Mar 00 issue)

\reference {} Burrows, A. \& Sharp, C.M. 1999. \apj, 512, 843

\reference {} Cuby, J.G., Saracco, P., Moorwood, A.F.M., D'Odorico, S.,
Lidman, C., Comeron, F., \& Spyromilio, J. 1999, \aap, 349 41 

\reference {} Golimowski, D.A. et al. 1998, \aj, 115, 2579 

\reference {} Griffith, C.A., Yelle, R.V., \& Marley, M.S. 1998 {\it
Science}, 282, 2063 

\reference {} Khare, B.N. \& Sagan, C. 1984, Icarus, 60, 127

\reference {} Kirkpatrick, J.D., Henry, T., \& McCarthy, D.W. 1991,
\apjs, 77, 419 

\reference {} Kirkpatrick, J.D., Reid, I.N., Liebert, J., Cutri, R.M., 
Nelson, B., Beichman, C.A., Dahn, C.C., Monet, D.G., Skrutskie, M.F., 
and Gizis, J. 1999, \apj, 519, 802 

\reference {} Leggett, S.K., Toomey, D.W., Geballe, T.R., \& Brown, R.H.
1999, \apjl, 517, L139

\reference {} Lodders, K. 1999, \apj, 519, 793

\reference {} Mart{\'{\i}}n, E.L., Delfosse, X., Basri, G., Goldman, B.,
Forveille, T., \& Zapatero Osorio, M.R. 1999, \aj, 118, 2466 

\reference {} Nakajima, T., Oppenheimer, B.R., Kulkarni, S.R.,
Golimbowski, D.A., Matthews, K. \& Durrance, S.T. 1995, {\it Nature},
{\bf 378}, 463

\reference {} Nakajima, T., Tsuji, T., Maihara, T., Iwamuro, F.,
Motohara, K., Taguchi, T., Hata, R. Tamura, M., \& Yamashita, T. 2000,
{\it PASJ}, in press. 

\reference {} Nefedov, A.P., Sinel'shchikov, V.A., \& Usachev, A.D.
1999, {\it Physica Scripta}, {\bf 59}, 432 

\reference {} Oppenheimer, B.R., Kulkarni, S.R., Matthews, K. \&
Nakajima, T., 1995, {\it Science}, {\bf 270}, 1478

\reference {} Oppenheimer, B.R., Kulkarni, S.R., Matthews, K. \& van
Kerkwijk, M.H. 1998, \apj, 502, 932

\reference {} Pavlenko, Ya., Zapatero Osorio, M.R., \& Rebolo, R. 2000,
{\it Astr. Ap.}, in press 

\reference {} Reid, I.N., Kirkpatrick, J.D., Williams, R., Liebert, J., 
\& Burgasser, 2000, {\it Astronomical Journal}, in press 

\reference {} Schultz, A.B. et al. 1998, \apjl, 492, L181

\reference {} Strauss, M.A., Fan, X., Gunn, J.E., Leggett, S.K.,
Geballe, T.R., Pier, J.R., Lupton, R.H., Knapp, G.R., et al. 1999, 
\apjl, 522, L61 (S99) 

\reference {} Tsuji, T., Ohnaka, K., \& Aoki, W. 1999, \apjl, 520, L119

\reference {} Tsvetanov, Z.I. et al. 2000, submitted to ApJ 

\end{references}
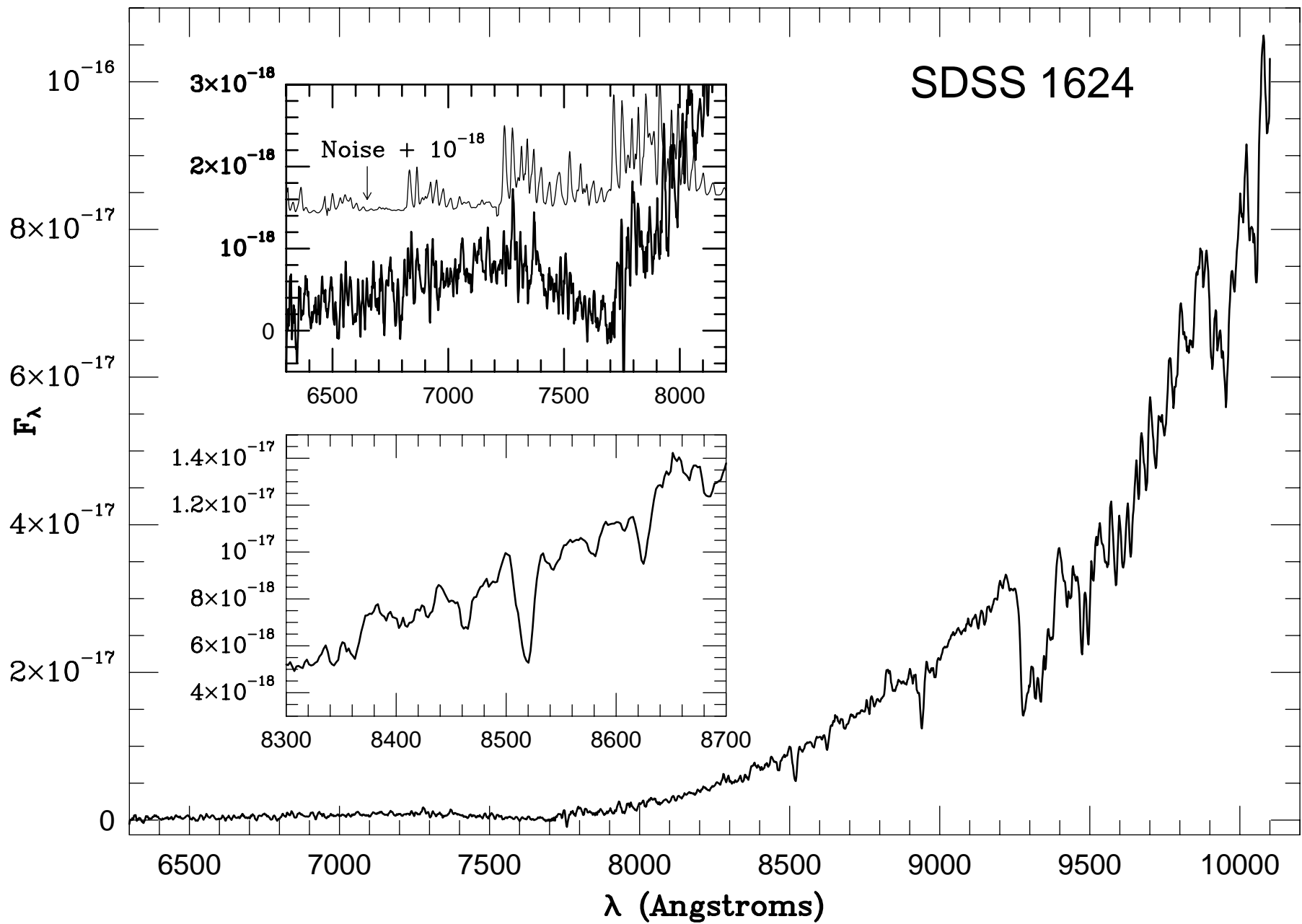
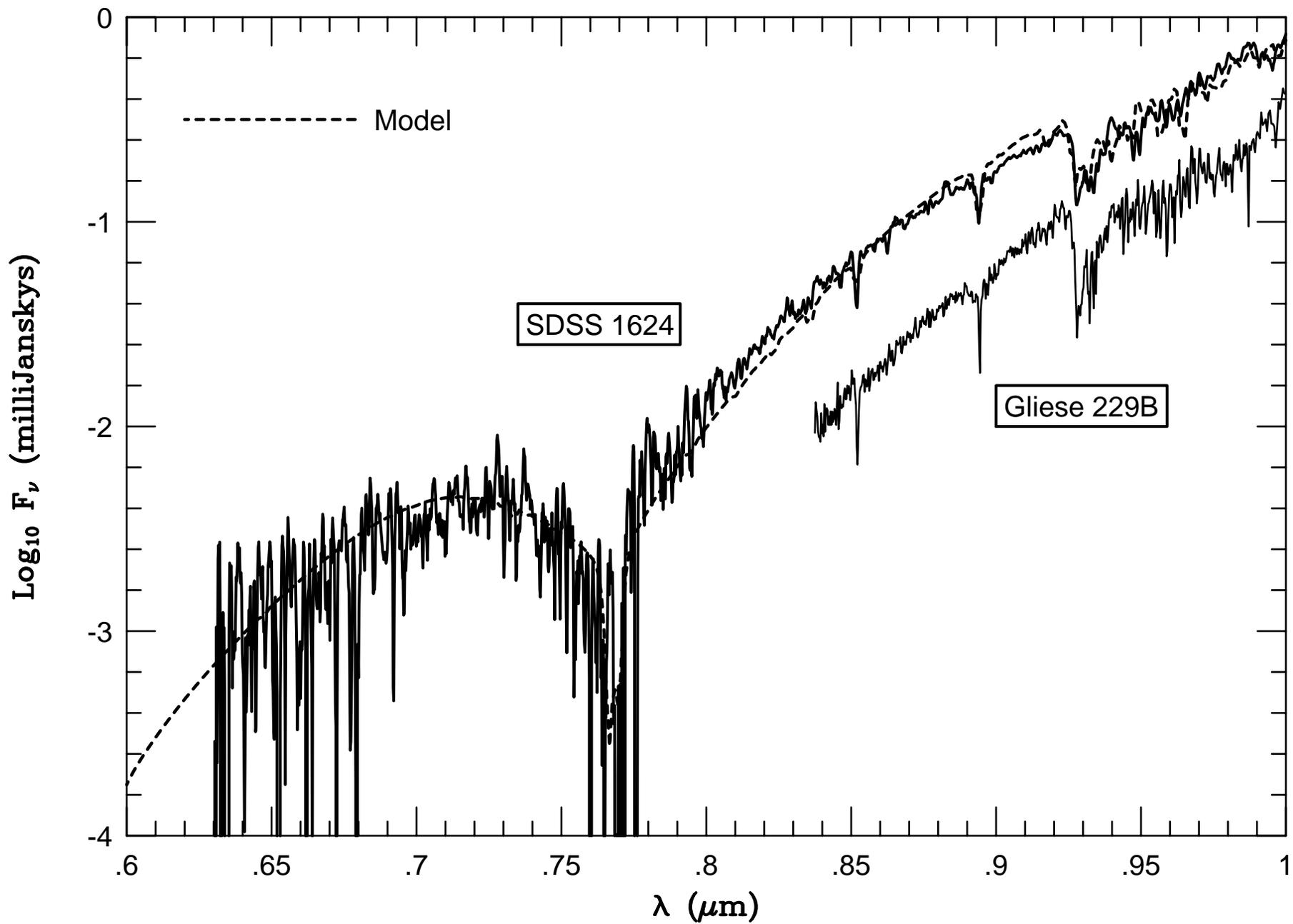
\end{document}